\documentstyle[twoside,fleqn,espcrc2]{article}
\newcommand{\half}{{\scriptstyle{{1\over 2}}}}
\def\beqa{\begin{eqnarray}}
\def\eeqa{\end{eqnarray}}
\def\bea{\begin{array}}
\def\eea{\end{array}}
\def\cT{{\beta}}

\def\cA{{\cal{A}}}
\def\acosh{{\rm acosh}} 
\def\diag{{\rm diag}} 
\def\tr{{\rm tr}} 
\def\Tr{{\rm Tr}} 
\def\pl{{{\cal P}_\infty}}
\def\plo{{{\cal P}_\infty^0}}

\title{\vskip-7mm 
Reflections on Topology - Viewpoints on Abelian Projections
\thanks{Presented at the workshops ``Lattice Hadron Physics", 9-18 July, 
Cairns, Australia and ``Light-cone Physics: Particles and Strings", 3-11 
September 2001, Trento, Italy.
}\vskip-2.3cm\hfill\small INLO-PUB-04/01\vskip2cm
}
\author{Pierre van Baal\address{Instituut-Lorentz for Theoretical Physics, 
University of Leiden,\\PO Box 9506, NL-2300 RA Leiden, The Netherlands.}}
\begin{document}
\begin{abstract}
This talk discusses two topological features in non-abelian gauge theories,
related by the notion of abelian projection and the Hopf invariant. 
Minimising the energy of the non-linear sigma model with a Skyrme-like term 
(the Faddeev-Niemi model), can be identified with a non-linear maximal abelian 
gauge fixing of the $SU(2)$ gauge vacua with a winding number equal to the 
Hopf invariant. In the context of abelian projection the Hopf invariant can 
also be associated to a monopole world line, through the Taubes winding, 
measuring its contribution to topological charge. Calorons with non-trivial 
holonomy provide an explicit realisation. We discuss the identification of 
its constituent monopoles through degenerate eigenvalues of the Polyakov 
loop (the singularities or defects of the abelian projection). It allows 
us to study the correlation between the defect locations and the explicit
constituent monopole structure, through a specific $SU(3)$ example.
\begin{center}
\vskip1mm
{\Large In memory of all those who lost their lifes on 11 September 2001}
\vskip-8mm
\end{center}
\end{abstract}
\maketitle
\section{Introduction}

Abelian projection was introduced by 't Hooft in an attempt, through a suitable 
choice of gauge, to decompose a non-abelian gauge field in its neutral and 
charged components~\cite{'tHooft:1981ht}. In its simplest form it involves 
choosing an observable $X(x)$ that transforms under gauge transformations
as $g^\dagger(x)X(x)g(x)$, which can be used to diagonalise $X(x)$. This 
can be done in a smooth way when none of the eigenvalues coincide. The 
remaining gauge freedom is $U(1)^r$, where $r$ is the rank of the gauge
group. These are associated with the $r$ neutral gauge bosons in this 
gauge.  Singularities occur when two (or more) eigenvalues coincide, and 
these can in three dimensions be shown to give rise to (generically) 
point-like singularities representing magnetic monopoles, as defined 
with respect to the remnant abelian gauge group. 

A smoother, but non-local, abelian gauge fixing can be 
introduced~\cite{'tHooft:1981ht} by taking an abelian field as a background 
(e.g. for $SU(2)$ the component proportional to $\tau_3=\diag(1,-1)$), and 
imposing the background gauge condition on the charged component of the gauge 
field. This can be formulated by minimising $\int |A_\mu^{\rm ch}(x)|^2$ 
along the gauge orbit.

Inspired by the abelian projection, Faddeev and Niemi attempted to identify 
the field $\vec n(x)$ (here of unit length, $|\vec n(x)|=1$) in an $O(3)$ 
non-linear sigma model with the local colour direction for $SU(2)$ gauge 
theory. The hope was that the static {\em knotted} solutions constructed 
numerically~\cite{Faddeev:1997zj} in the model originally introduced by 
Faddeev~\cite{Faddeev:1975tz}, were possibly related to 
glueballs~\cite{Faddeev:1999eq}. A difficulty is that one would not expect 
dynamics for a quantity that is associated with the colour direction, due 
to gauge invariance. Furthermore, the $O(3)$ symmetry is in general 
spontaneously broken for non-linear sigma models in 3+1 dimensions. The 
associated Goldstone bosons are unwanted in non-abelian gauge theories 
without matter. We will see that nevertheless the $\vec n$ field {\em can} be 
identified with an $SU(2)$ gauge field, albeit with {\rm zero} field 
strength. Under this identification we have $(\partial_\mu\vec n(x))^2=
4|A_\mu^{\rm ch}(x)|^2$, giving a hint that minimising the $O(3)$ energy 
functional can be interpreted~\cite{vanBaal:2001jm} in terms of maximal 
abelian gauge fixing, as will be discussed in section 2. 

The static knotted solutions, as maps $\vec n(\vec x)$ from $S^3$ (compactified 
$R^3$) to $S^2$, are classified by the Hopf invariant. The pre-image of a 
generic value of $\vec n$ traces out a loop in $R^3$ (i.e. the collection of 
points $\vec x$, where $\vec n(\vec x)=\vec n$). The linking number of any 
two of such loops is equal to this Hopf invariant. This also coincides with 
the winding number of the gauge function $g(\vec x)$, such that 
$n^a(\vec x)\tau_a=g(\vec x)\tau_3 g^\dagger(\vec x)$, with $\tau_a$ the 
Pauli matrices. The associated $SU(2)$ gauge field with zero field strength 
is $A(\vec x)=g^\dagger(\vec x)dg(\vec x)$.

Such a relation between Hopf invariant and topological charge can also occur 
for monopoles. A basic monopole is characterised by a hedgehog, with the Higgs 
field defining the colour direction, pointing out radially. If we gauge rotate 
the monopole (for the spherically symmetric case, equivalent to a real rotation)
while moving along its worldline, a particular point of the hedgehog will trace
out a loop whose linking with the monopole worldline is an invariant. This was
used by Taubes to make from monopole fields, configurations with non-zero 
topological charge~\cite{Tau}. In his formulation a monopole-antimonopole 
pair is created and separated to a finite distance. Kept at this separation 
one of them is rotated around the axis connecting the two. After this is 
completed they are brought together and made to annihilate. The time is 
considered to be euclidean and the rotation introduces a ``twist" in the 
field that prevents the four dimensional configuration from decaying to 
$A=0$. The obstruction is precisely the topological charge, whose value 
is equal to the net number of rotations. 

\begin{figure}[htb]
\vspace{4cm}
\includegraphics{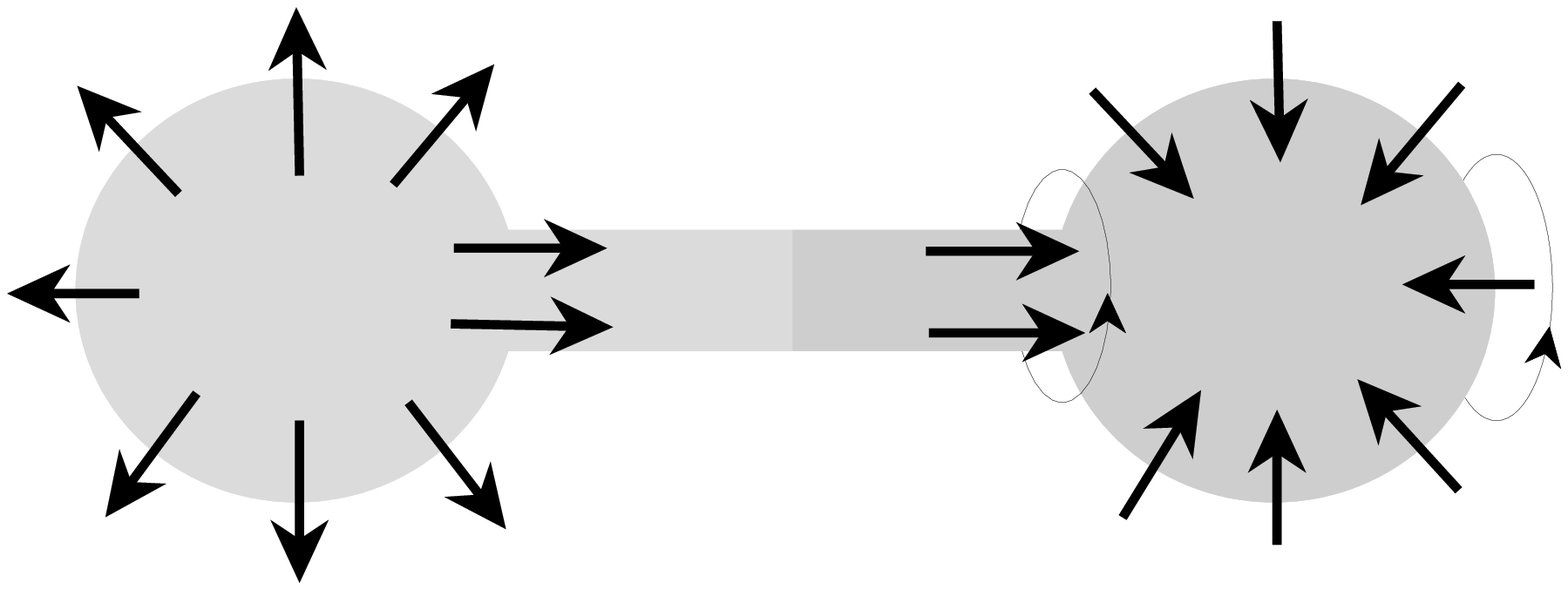}
\includegraphics{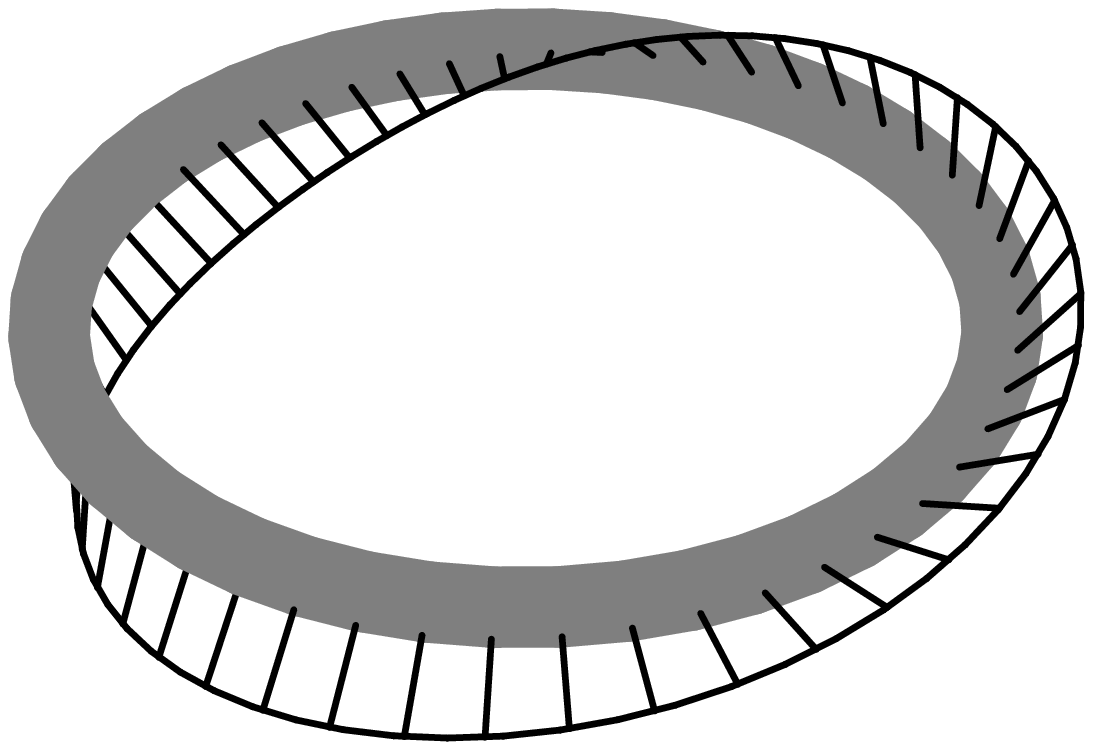}
\caption{Topological charge constructed from oppositely charged
monopoles by rotating one of them. For a closed monopole line, the 
embedding of the unbroken subgroup makes a full rotation.}
\label{fig:taubes}
\end{figure}

This process can just as well be described in terms of a closed monopole 
loop (in the same way that the Wilson loop is associated with the creation, 
propagation and annihilation of a heavy quark-antiquark pair), see 
Fig.~\ref{fig:taubes}. The identification with the Hopf invariant should be 
understood in the following sense: the orientation of the monopole is 
described by $SU(2)/U(1)\!\sim\!S^2$ at each point on the monopole loop 
($S^1$), describing a twisted $S^2$ bundle over $S^1$, e.g. making the total 
space into an $S^3$ for one full ``frame" rotation. This is the Hopf fibration,
although we have interchanged fibre and base space as compared to the usual 
formulation.

At finite temperature $A_0$ plays the role of the Higgs field and the calorons 
(periodic instantons) provide an explicit example of this Taubes winding, when
it has a non-trivial value of the Polyakov loop at infinity. This Polyakov loop
is independent of the directions at infinity, because the finite action of the 
caloron forces the field strength to vanish at infinity. In this case the 
caloron splits in ($n$ for $SU(n)$) constituent monopoles, which are the basic 
spherically symmetric BPS monopoles. Of these, $n-1$ are time independent, 
whereas the time dependence of the other exactly coincides with the rotation 
(at a  uniform rate)~\cite{KvB2,KvBn} of the Taubes winding. In terms of the 
abelian projection and the introduction of a composite Higgs field the relation
of the Hopf invariant with twisted monopole loops has been extensively studied 
by Jahn~\cite{Jahn}.

In section 3 we will use the Polyakov loop as the observable $X(\vec x)$ to 
be diagonalised, a version of abelian projection that is non-local in time. 
For the application to the calorons this is not a problem, since at high 
temperature the action density is static. The static action density is 
not in conflict with the Taubes winding, as this is associated to a gauge 
rotation.  Defects occur when eigenvalues coincide. Their location is not 
an artifact of the gauge, since eigenvalues are gauge invariant. But another 
choice of $X$ will give generically defects at other locations. In the 
light of this there seems no reason to expect the defects to always be
associated to physical lumps. We study in how far this is true in the 
case where two of the constituents for the $SU(3)$ caloron are brought 
close together~\cite{VanBaal:2001pm}.

In section 4 we end with some conclusions concerning the role of topology
in non-abelian gauge theories.

\section{The Faddeev-Niemi model}

The model is defined in terms of a three-vector $\vec n(x)$ of fixed (here
chosen unit) length. To allow for non-trivial static solutions a Skyrme-like
higher-order term is added~\cite{Faddeev:1975tz}, through the introduction
of a composite gauge field strength $F_{\mu\nu}(x)=\frac{1}{2}\vec n(x)\cdot
(\partial_\mu\vec n(x)\wedge\partial_\nu\vec n(x))$.
A useful identity is $F^2_{\mu\nu}(x)=\frac{1}{4}(\partial_\mu\vec n(x)\wedge
\partial_\nu\vec n(x))^2$, which follows from the fact that
$\partial_\mu\vec n(x)\wedge\partial_\nu\vec n(x)=2F_{\mu\nu}(x)\vec n(x)$.
Its proportionality to $\vec n(x)$ follows from the fact that the latter is
perpendicular to $\partial_\mu\vec n(x)$ (since $\vec n(x)$ is a unit
three-vector). The action is given up to an overall factor by
\begin{equation}
S\!=\!\!\int\!\! d^4x\!\left(\!\partial_\mu\vec n(x)\cdot\partial^\mu
\vec n(x)\!-\!\frac{1}{2}F_{\mu\nu}(x)F^{\mu\nu}(x)\!\right)\!,\label{eq:O3}
\end{equation}
brought to this simple form by a suitable rescaling of $x$. Finite energy 
requires $\vec n(\vec x)$ to approach a constant vector at spatial infinity. 
In this way static configurations are classified by the topological maps from 
$S^3$ into $S^2$, characterised by the Hopf invariant. The two-form $F(\vec x)
=\vec n(\vec x)\cdot(d\vec n(\vec x)\wedge d\vec n(\vec x))$ implicitly 
defines an abelian gauge field one-form $A(\vec x)$ through $F(\vec x)=
dA(\vec x)$, in terms of which the Hopf invariant is given by 
$Q=\frac{1}{4\pi^2}\int A(\vec x)\wedge F(\vec x)$. Remarkably, the 
energy is bounded by a fractional power of this Hopf 
invariant~\cite{Vakulenko:1979uw,Kundu:1982bc}.
\begin{equation}
E=\int d^3x\left((\partial_i\vec n(x))^2+\frac{1}{2}F_{ij}^2(x)\right)
\geq c|Q|^{3/4},\label{eq:EO3}
\end{equation}
with $c=16\pi^2 3^{3/8}$. This gives a rough bound, which can be improved 
on~\cite{Ward:1998pj} (by roughly a factor 2). Extensive numerical 
studies~\cite{Battye:1998pe,Hietarinta:2000ci} have gone up to $Q=8$, 
with energies indeed following the fractional power of $Q$.

\subsection{Reformulations}

We reformulate the non-linear sigma model in two steps, both well 
known~\cite{Eichenherr:1978qa,D'Adda:1978uc}. The first involves $CP_1$ 
fields. Its main advantage is that the abelian gauge field appearing in the 
Hopf invariant, no longer needs to be defined implicitly. To be specific, one 
introduces a complex two-component field $\Psi(x)$, also having unit length.
A further phase is removed by the local abelian gauge invariance, obvious 
from the following relation to the $n$ field:
\begin{equation}
n^a(x)=\Psi^\dagger(x)\tau^a\Psi(x).
\end{equation}
The abelian gauge invariance of the $CP_1$ model leads to a composite 
gauge field
\begin{equation}
A_\mu(x)=-i\Psi^\dagger(x)\partial_\mu\Psi(x),
\end{equation}
and one verifies that indeed $F(x)=dA(x)$. Useful identities for these 
computations are the completeness relation $\delta_{ij}\delta_{kl}+
\tau^a_{ij}\tau^a_{kl}=2\delta_{il}\delta_{jk}$ and $i\varepsilon_{abc}
\tau^b_{ij}\tau^c_{kl}=\tau^a_{kj}\delta_{il}-\tau^a_{il}\delta_{jk}$. 
The energy, Eq.~(\ref{eq:EO3}), becomes
\begin{equation}
E=\!\int d^3x\left(4|D_i\Psi(\vec x)|^2-
\frac{1}{2}F_{ij}^2(\vec x)\right),\label{eq:CP1}
\end{equation}
with $D_\mu\!=\!\partial_\mu\!-\!iA_\mu(x)$ the covariant derivative. 

Next we make use of the fact that any complex two-component vector of unit 
length is in one to one relation to an $SU(2)$ group element. Alternatively 
we can write $\Psi(x)=g(x)\Psi_0$. For convenience we choose $\Psi^\dagger_0=
(1,0)$, such that
\begin{equation}
n_a(x)=\frac{1}{2}{\rm tr}\left(\tau_3 g^\dagger(x)\tau_a g(x)\right). 
\end{equation}
We introduce $J_\mu(x)\equiv i\tau_a J_\mu^a(x)\equiv g^\dagger(x)\partial_\mu 
g(x)$, which can be interpreted as the components of an $SU(2)$ gauge 
connection with vanishing curvature, that is $G(x)\!=\!dJ(x)+J(x)\!\wedge\!J(x)
\!=\!0$, where $J(x)\!\equiv \!J_\mu(x)dx_\mu$. A simple calculation shows that 
$A_\mu(x)\!=\!J_\mu^3(x)$ and $\partial_\mu\Psi^\dagger(x)\partial^\mu\Psi(x)
\!=\!J_\mu^a(x)J_a^\mu(x)$. The zero non-abelian field strength leads to 
$F_{\mu\nu}(x)=2(J_\mu^1(x)J^2_\nu(x)-J_\nu^1(x)J^2_\mu(x))$, or 
\begin{equation}
F(x)=dJ^3(x)=2J^1(x)\wedge J^2(x),
\end{equation}
in terms of $J^a(x)\equiv J^a_\mu(x)dx_\mu$.
This implies that $A(\vec x)\wedge F(\vec x)=2J^3(\vec x)\wedge J^1(\vec x)
\wedge J^2(\vec x)$, and substituting $J(x)=g^\dagger(x)dg(x)$ one finds 
that the Hopf invariant is exactly equal~\cite{Battye:1998pe} to the winding 
number of the gauge function $g(\vec x)$, 
\begin{equation}
\frac{1}{4\pi^2}A(\vec x)\wedge F(\vec x)=
\frac{1}{24\pi^2}{\rm tr}(g^\dagger(\vec x)dg(\vec x))^3.
\end{equation}
Using that $J$ has zero curvature, we can of course also relate the Hopf
invariant to the non-abelian Chern-Simons form, giving 
$\frac{1}{4\pi^2}A(\vec x)\wedge F(\vec x)=-\frac{1}{8\pi^2}{\rm tr}[J(\vec x)
\wedge\!\left(dJ(\vec x)+\frac{2}{3}J(\vec x)\wedge\!J(\vec x)\right)]
\equiv\!CS(J)$. 

\subsection{The gauge fixing interpretation}

The natural question to ask now, is what the interpretation of the energy
functional becomes in terms of the non-abelian gauge field 
\begin{equation}
E=\!\!\int\!\! d^3x\!\left(\!4\left(J^\alpha_i(\vec x)\right)^2\!\!+\!2\left(
\epsilon_{\alpha\beta}J^\alpha_i(\vec x)J^\beta_j(\vec x)\right)^2\right),
\end{equation}
where the index $\alpha$ and $\beta$ run only over 1 and 2. With 
the absence of the neutral component, $J_\mu^3(x)$, the $SU(2)/U(1)$ 
formulation is evident. It defines a positive definite functional on 
the gauge orbit, and its minimum can thus be seen as a particular 
(non-linear) maximal abelian gauge fixing, leaving the abelian subgroup 
generated by $\tau_3$ unfixed~\cite{'tHooft:1981ht,Kronfeld:1987ri}. 
As the reparametrisation is mathematically equivalent, we are entitled to 
interpret the minima of the energy functional in the sector with a given 
value of $Q$ as gauge fixed pure gauge (i.e. curvature free, or flat) 
connections in a sector with gauge field winding number $Q$. Therefore, there 
is a gauge fixing in terms of which the gauge vacua with different winding 
number can be characterised by inequivalent knots. We thus conclude that 
instantons ``knit", interpolating between different types of ``knots".

The relation between $\vec n$ and flat $SU(2)$ gauge fields can also be 
understood from geometric quantisation on $SU(2)/U(1)\!\sim\!S^2$ and 
chiral models, but the relation to gauge fixing was only noted 
recently~\cite{vanBaal:2001jm}. Finding the {\em absolute} minimum of a gauge 
fixing functional is known to be a hard problem, seemingly reflected in the 
difficulty of finding the knots with the lowest energy for a given value of 
$Q$, as clearly illustrated in Fig.~2 of Ref.~\cite{Hietarinta:2000ci}

\subsection{Implications}

Much work has been invested in interpreting this model as an effective 
low-energy representation of $SU(2)$ gauge 
theory~\cite{Faddeev:1999eq,Langmann:1999nn,Cho:1999wp}. The quality 
of this approximation is being investigating by inverse Monte Carlo 
techniques~\cite{Dit}. Amongst other things, these numerical results 
have reported the presence of the unwanted massless Goldstone bosons, 
if one does not include terms that explicitly break the $O(3)$ symmetry.
Also a perturbative renormalisation group study showed there are 
difficulties with the restricted formulation of the model, as an effective 
description of $SU(2)$ gauge theory~\cite{Gies:2001hk}. As we will argue,
this difficulty seems to be related to the fact that one expands around 
a background that is actually unstable. 

The conventional starting point in these studies is based on the 
decomposition~\cite{Cho:1980nv}
\begin{equation}
\vec A_\mu(x)\!=\!\partial_\mu\vec n(x)\wedge\vec n(x)+C_\mu(x)\vec n(x)+
\vec W_\mu(x),
\end{equation}
where $C_\mu$ describes an abelian gauge field and $W^a_\mu$ are the charged 
components of the field, with respect to $n^a(x)\tau_a$. One attempts to 
integrate out both $W^a_\mu$ {\em and}, unlike in abelian projection, 
$C_\mu$.  In perturbation theory this amounts to a background field 
calculation. For later use we consider the one parameter family of backgrounds 
$\vec A_\mu(x)\!=\!q\partial_\mu\vec n(x)\wedge\vec n(x)$. For $q\!=\!1$ 
$\vec n(x)$ is covariantly constant, i.e. $\partial_\mu\vec n(x)+\vec A_\mu(x) 
\wedge\vec n(x)\!=\!0$, which remains true when adding $C_\mu\vec n$ to the 
background. It ensures that the abelian gauge field transforms properly 
under gauge transformations after abelian projection~\cite{Cho:1980nv}.
To address the issue of perturbative stability, we compute the field strength 
for general $q$,
\begin{equation}
\vec F_{\mu\nu}(\vec x)=q(q-2)\partial_\mu\vec n(\vec x)\wedge\partial_\nu
\vec n(x),
\end{equation}
which is non-zero for $q\!=\!1$. The energy of this background field is 
proportional to $q^2(2-q)^2$ and actually has a local maximum at $q\!=\!1$, 
and hence is unstable. On the other hand, the energy of the background 
vanishes not only for $q\!=\!0$ (the trivial background), but also for 
$q\!=\!2$. The latter corresponds to the flat connection $J$ that has 
topological charge equal to $Q$, given by the integral of the non-abelian 
Chern-Simons form over space. Rescaling the flat connection $J$ with $q/2$ one 
easily finds its value for any $q$, $\int CS(qJ(\vec x)/2)\!=\!q^2(3-q)Q/4$. 
For $q\!=\!1$, this relation between the Hopf invariant and the non-abelian 
Chern-Simons form can also be found in Ref.~\cite{Langmann:1999nn}. We note 
that the integral gives half the Hopf invariant for $q\!=\!1$, thus in some 
sense this background is ``half-way" between two vacua.

For $q=2$ the (flat) background is useful in separating the gauge field in its 
different topological sectors, since small fluctuations parametrised by 
$C_\mu$ and $W^a_\mu$ do not change the value of the winding number. With 
what we could call the Faddeev-Niemi gauge fixing, such a separation in 
different topological sectors is now possible in {\em localised} form, without
having to appeal to the {\em asymptotics} of the gauge field. This may lead 
to new tools and insights to deal with the non-trivial topology of non-abelian 
gauge theories.

\section{The SU(3) Caloron}
We now turn our attention to finite temperature instantons (calorons) with a
Polyakov loop, $P(\vec x)\!=\!P\exp(\int_0^\cT\! A_0(\vec x,t) dt)$ in the 
periodic gauge $A_\mu(t,\vec x)\!=\!A_\mu(t+\cT,\vec x)$, non-trivial at 
spatial infinity (specifying the holonomy). It implies the spontaneous 
breakdown of gauge symmetry. For a charge one $SU(n)$ caloron, their are
$n$ constituent monopoles~\cite{KvBn}. Locations of constituents monopoles 
can be identified through: \\
$\bullet$ Points where two eigenvalues of the Polyakov loop coincide, which is
where the $U^{n\!-\!1}(1)$ symmetry is partially restored to 
$SU(2)\!\times\!U^{n\!-\!2}(1)$.\\
$\bullet$ The centers of mass of the (spherical) lumps. \\
$\bullet$ The Dirac monopoles (or rather dyons, due to self-duality) as the 
sources of the abelian field lines, extrapolated back to the cores.\\ If well 
separated and localised, all these coincide~\cite{KvB2,Dub}. This is no longer 
the case when two constituents come close together, as shown for $SU(2)$ in 
Ref.~\cite{GGMvB}, and for $SU(3)$ in Ref.~\cite{VanBaal:2001pm}.

The locations of the constituent monopoles may be chosen at will, but their
masses are fixed by the eigenvalues of $\pl\!\equiv\!\lim_{|\vec x|\rightarrow
\infty}P(\vec x)$. These eigenvalues can be ordered by a constant gauge 
transformation $W_\infty$
\beqa
&&\hskip-3mm W_\infty^\dagger\pl W_\infty\!=\!\plo\!=\!\exp[2\pi i\,\diag
(\mu_1,\ldots,\mu_n)],\nonumber\\ &&\hskip-3mm\mu_1\leq\ldots\leq\mu_n\leq
\mu_{n+1}\!\equiv\!1\!+\!\mu_1,\label{eq:hol}
\eeqa
with $\sum_{m=1}^n\mu_m\!=\!0$. Using the classical scale invariance to put 
the extent of the euclidean time direction to one, $\beta=1$, the masses
of the constituent monopoles are now given by $8\pi^2\nu_i$, where 
$\nu_i\equiv\mu_{i+1}-\mu_i$. 

Similarly we can bring $P(\vec x)$ to the diagonal form, with eigenvalues 
ordered on the circle, by a {\em local} gauge function, $P(\vec x)=W(\vec x)
P_0(\vec x)W^\dagger(\vec x)$. We note that $W(\vec x)$ (unique up to a 
residual abelian gauge rotation) and $P_0(\vec x)$ will be smooth, except 
where two (or more) eigenvalues coincide. The ordering shows there are $n$ 
different types of singularities (called defects~\cite{FTW}), for each of 
the {\em neighbouring} eigenvalues to coincide. The first $n-\!1$ are 
associated with the basic monopoles (as part of the inequivalent $SU(2)$ 
subgroups related to the generators of the Cartan subgroup). The $n^{\rm th}$ 
defect arises when the first and the last eigenvalue (still neighbours on the 
circle) coincide. Its magnetic charge ensures charge neutrality of the 
caloron, and it carries the Taubes winding supporting the non-zero topological 
charge~\cite{KvB2,KvBn,CKvB}.

The topological charge can be reduced to surface integrals near the 
singularities with the use of $\tr(P^\dagger dP)^3\!=d~3 \tr((P_0^\dagger 
A_W P_0+2P_0^\dagger dP_0)\wedge A_W)=d~3\tr(A_W\!\wedge(2A_W\log P_0+P_0A_W 
P_0^\dagger))$, where $A_W\!\equiv W^\dagger dW$. If one assumes {\em all} 
defects are pointlike, this can be used to show that for each of the $n$ 
types the (net) number of defects has to equal the topological charge, the 
type being selected by the branch of the logarithm (associated with the $n$ 
elements in the center)~\cite{FTW}. This is the generic situation, but in 
special cases defects may form a submanifold, as we will find for a global 
embedding of the $SU(2)$ caloron in $SU(3)$.

\subsection{Coinciding constituents}

One might expect defects to merge when constituent monopoles do. The resulting 
triple degeneracy of eigenvalues for $SU(3)$ implies that the Polyakov loop 
takes on a value in the center. Yet this can be shown {\em not} to occur for 
the $SU(3)$ caloron with {\em unequal} masses. We therefore seem to have (at 
least) one more defect than the number of constituents, when two merging 
constituents will manifest themselves as one constituent. To study what 
happens in this case we first recall the simple formula for the $SU(n)$ 
action density~\cite{KvBn}
\beqa
&&\hskip-6mm\Tr F_{\mu\nu}^{\,2}(x)\!=\!\partial_\mu^2\partial_\nu^2\log\left[
\half\tr(\cA_n\cdots \cA_1)-\cos(2\pi t)\right],\nonumber\\
&&\hskip-6mm\cA_m\equiv\frac{1}{r_m}\left(\!\!\!\bea{cc}r_m\!\!&|\vec y_m\!\!
-\!\vec y_{m+1}|\\0\!\!&r_{m+1}\eea\!\!\!\right)\left(\!\!\!
\bea{cc}c_m\!\!&s_m\\s_m\!\!&c_m\eea\!\!\!\right),\label{eq:act}
\eeqa
with $\vec y_m$ the center of mass location of the $m^{\rm th}$ constituent 
monopole. We defined $r_m\!\equiv\!|\vec x\!-\!\vec y_m|$, $c_m\!\equiv\!
\cosh(2\pi\nu_m r_m)$, $s_m\!\equiv\!\sinh(2\pi\nu_m r_m)$, as well as
$\vec y_{n+1}\!\equiv\!\vec y_1$, $r_{n+1}\!\equiv\!r_1$. We are interested 
in the case where the problem of two coinciding constituents in $SU(n)$ is 
mapped to the $SU(n\!-\!1)$ caloron. For this we restrict to the case where
$\vec y_m\!=\!\vec y_{m+1}$ for some $m$, which for $SU(3)$ is {\em always} 
the case when two constituents coincide. Since now $r_m\!=\!r_{m+1}$, one 
easily verifies that $\cA_{m+1}\cA_m\!\!=\!\cA_{m+\!1}[\nu_{m+\!1}\!
\rightarrow\!\nu_m\!+\!\nu_{m+\!1}]$, describing a {\em single} constituent 
monopole (with properly combined mass), reducing Eq.~(\ref{eq:act}) to the 
$SU(n\!-\!1)$ action density, with $n\!-\!1$ constituents. 

\begin{figure}[htb]
\vspace{3cm}
\includegraphics{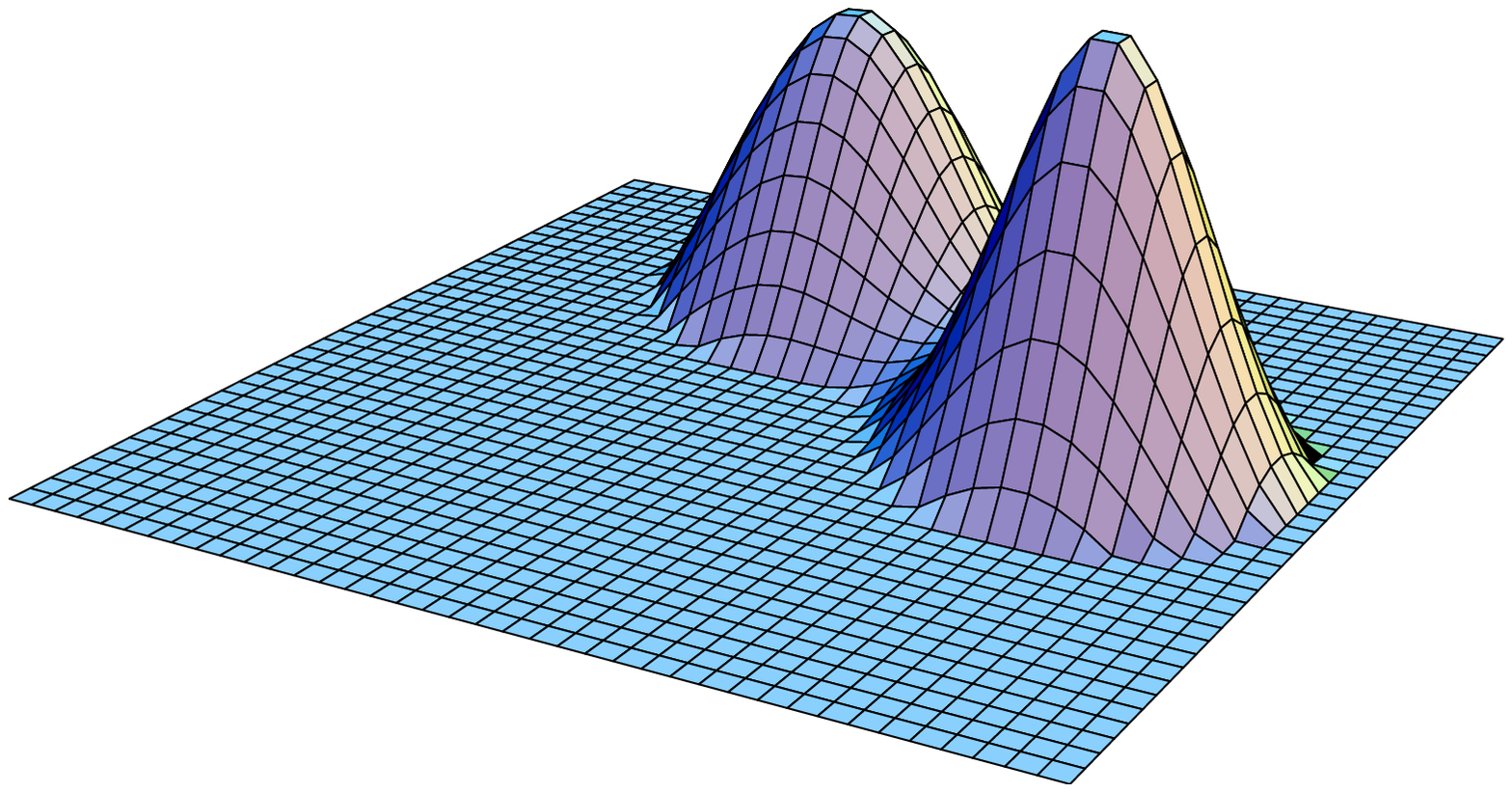}
\includegraphics{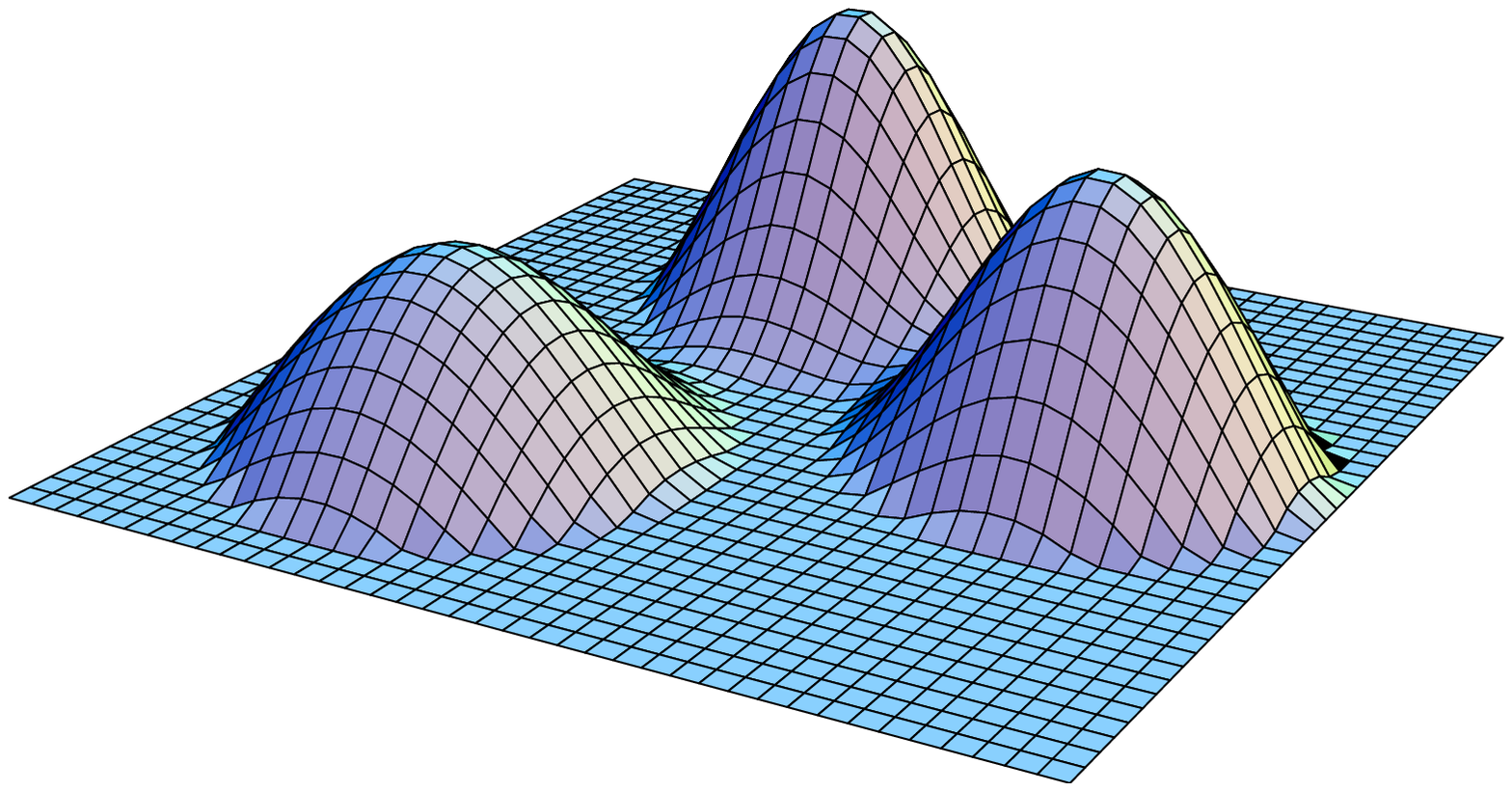}
\caption{Action densities for $SU(3)$ calorons on equal logarithmic scales,
cut off at $\half e^{-1}$, for $\beta\!=\!1$ in the plane defined by the 
constituent locations $\vec y_1\!=\!(-2,-2,0)$, $\vec y_2=(0,2,0)$, $\vec y_3
=(2,-1,0)$ and $t\!=\!0$, using $(\mu_1,\mu_2,\mu_3)=(-17,-2,19)/60$.
On the right we moved the first constituent to the location of the second.}
\label{fig:prof}
\end{figure}

\subsection{Tracing the defects}

We will study in detail a generic example in $SU(3)$, with $(\mu_1,\mu_2,\mu_3)
\!=\!(-17,-2,19)/60$. Fig.~\ref{fig:prof} shows the action density both for 
non-coinciding constituents and when moving the first constituent to the 
location of the second. We denote by $\vec z_m$ the position associated with 
the $m^{\rm th}$ constituent, where two eigenvalues of the Polyakov loop 
coincide (the defect locations). Numerically it was established 
that~\cite{Dub}, in the gauge where $\pl=\plo$ (see Eq.~(\ref{eq:hol})),
\beqa
P_1\!\!=\!\!P(\vec z_1)\!\!=\!\!\diag(\hphantom{-}e^{-\pi i\mu_3},\hphantom{-}
                    e^{-\pi i\mu_3},\hphantom{-}e^{2\pi i\mu_3}),\nonumber\\
P_2\!\!=\!\!P(\vec z_2)\!\!=\!\!\diag(\hphantom{-}e^{2\pi i\mu_1},\hphantom{-}
                    e^{-\pi i\mu_1},\hphantom{-}e^{-\pi i\mu_1}),\\
P_3\!\!=\!\!P(\vec z_3)\!\!=\!\!\diag(-e^{-\pi i\mu_2},\hphantom{-}
                    e^{2\pi i\mu_2},-e^{-\pi i\mu_2}).\nonumber
\eeqa
This is for {\em any} choice of holonomy and constituent locations (with the 
proviso they are well separated, i.e. their cores do not overlap, in which 
case to a good approximation $\vec z_m\!=\!\vec y_m$). Now we take $\vec y_1
\!=\!(0,0,10+d)$, $\vec y_2\!=\!(0,0,10\!-\!d)$ and $\vec y_3\!=\!(0,0,-10)$. 
The limit of coinciding constituents is achieved by $d\!\rightarrow\!0$. In 
very good approximation, as long as the first two constituents remain well 
separated from the third constituent (carrying the Taubes winding), $P_3$ 
will be constant in $d$ and the $SU(3)$ gauge field~\cite{Dub} of the first 
two constituents will be constant in time (in the periodic gauge). Thus 
$P(\vec z_m)=\exp(A_0(\vec z_m))$ for $m=1,2$. This simplifies the calculation 
of the Polyakov loop considerably.

\begin{figure}[htb]
\vspace{5.0cm}
\includegraphics{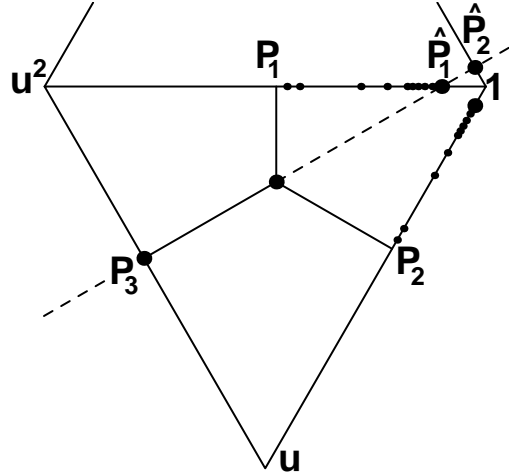}
\caption{The fundamental Weyl chamber with the positions of $P_m$ indicated 
at $d=2,$ 1, .2, .1, .05, .04, .03, .02, .01, .005, .001, .0005, and (large 
dots) 0. The perpendiculars point to $\pl$ (center), and emanate from the 
values of $P_m$ for well separated constituents. The dashed line shows the 
$SU(2)$ embedding for $d=0$. ($u\equiv\exp(2\pi i/3)\,{\bf 1}$)}
\label{fig:weyl}
\end{figure}

When the cores of the two approaching constituents start to overlap, $P_1$ 
and $P_2$ are no longer diagonal (mixing the lower $2\!\times\!2$ components). 
At $d=0$ they are diagonal again, but $P_2$ will be no longer in the 
fundamental Weyl chamber (the ``logarithm" of the Cartan subgroup). A Weyl 
reflection maps it back, while for $d\neq0$ a more general gauge rotation 
back to the Cartan subgroup is required to do so, see Fig.~\ref{fig:weyl}. For 
$d\!\rightarrow\!0$, {\em each} $P_m$ (and $\pl$) falls on the dashed line, 
defining the reduction to $SU(2)$. 

\begin{figure}[htb]
\vspace{4.6cm}
\includegraphics{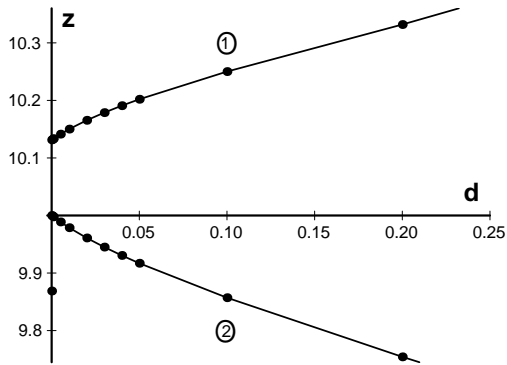}
\caption{The defect locations $\vec z_1$ and $\vec z_2$, along the $z$-axis,
for $M\!\equiv\!\nu_2\!-\!\nu_1\!=\!0.1$ as a function of $d$.}
\label{fig:trace}
\end{figure}

To illustrate this more clearly, we give the expressions for $P_m$ (which we 
believe to hold for any non-degenerate choice of the $\mu_i$) when 
$d\rightarrow0$:
\beqa
\hat P_1\!\!=\!\!P(\vec z_1)\!\!=\!\!\diag(\hphantom{-}e^{2\pi i\mu_2},
        \hphantom{-}e^{2\pi i\mu_2},\hphantom{-}e^{-4\pi i\mu_2}),\!\nonumber\\
\hat P_2\!\!=\!\!P(\vec z_2)\!\!=\!\!\diag(\hphantom{-}e^{-\pi i\mu_2},
         \hphantom{-}e^{2\pi i\mu_2},\hphantom{-}e^{-\pi i\mu_2}),\!\\
\hat P_3\!\!=\!\!P(\vec z_3)\!\!=\!\!\diag(-e^{-\pi i\mu_2},
          \hphantom{-}e^{2\pi i\mu_2},-e^{-\pi i\mu_2}).\!\nonumber
\eeqa
These can be factorised as $\hat P_m=\hat P_2 Q_m$, where $\hat P_2$ describes 
an overall $U(1)$ factor. In terms of $Q_1\!=\!\diag(e^{3\pi i\mu_2},1, 
e^{-3\pi i\mu_2})$, $Q_2\!=\!\diag(1,\!1,\!1)\!=\!{\bf 1}$ and $Q_3\!=
\!\diag(-1,1,-1)$ the $SU(2)$ embedding in $SU(3)$ becomes obvious. It 
leads for $Q_2$ to the trivial and for $Q_3$ to the non-trivial element 
of the center of $SU(2)$ (appropriate for the latter, carrying the Taubes 
winding). On the other hand, $Q_1$ corresponds to $\diag(e^{3\pi i\mu_2},
e^{-3\pi i\mu_2})$, which for the $SU(2)$ caloron is not related to coinciding 
eigenvalues. For $d\!\rightarrow\!0$, Fig.~\ref{fig:trace} shows that $\vec 
z_1$ gets ``stuck'' at a {\em finite} distance (0.131419) from $\vec z_2$.

\subsection{Spurious defects}

The $SU(2)$ embedding determines the caloron solution for $d\!=\!0$, with 
constituent locations $\vec y_1^{\,\prime}\!=\!\vec y_2$ and $\vec y_2^{\,
\prime}\!=\!\vec y_3$, and masses $\nu_1^{\,\prime}\!=\!\nu_1\!+\nu_2\!=\!
\mu_3\!-\!\mu_1$ and $\nu_2^{\,\prime}\!=\!\nu_3$. The spurious nature of 
the defect is obvious by calculating its location purely in terms of this 
$SU(2)$ caloron, demanding the $SU(2)$ Polyakov loop to equal $\diag(e^{3\pi 
i\mu_2},e^{-3\pi i\mu_2})$. For this we can use the analytic 
expression~\cite{GGMvB} of the $SU(2)$ Polyakov loop along the 
$z$-axis. The location of the spurious defect, $\vec z_1\!=\!(0,0,z)$, is 
found by solving $3\pi\mu_2\!=\!\pi\nu_2^{\,\prime}\!-\!\half\partial_z\acosh[
\half\tr(\cA_2^{\prime}\cA_1^{\prime})]$. For our example, $z\!=\!10.131419$ 
indeed verifies this equation. 

\begin{figure}[b]
\vspace{3.3cm}
\includegraphics{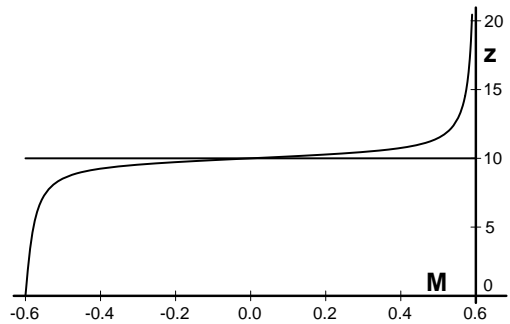}
\caption{The defect locations $\vec z_1$ and $\vec z_2$, along the $z$-axis, as
a function of $M\!\equiv\!\nu_2\!-\!\nu_1$ for $d\!\rightarrow\!0$.}
\label{fig:gen}
\end{figure}

Fig.~\ref{fig:gen} gives the spurious location as a function of the mass 
difference of the two coinciding constituents. We find that only when this 
difference {\em vanishes}, the defects merge to form a triple degeneracy. Using 
the relation $3\mu_2\!=\!\nu_1\!-\!\nu_2$, the case of {\em equal} masses for 
the coinciding constituents corresponds to $\mu_2\!=\!0$. For {\em unequal} 
masses the defect is always spurious, but it tends to stay within reach of the 
non-abelian core of the coinciding constituent monopoles, 
$(\pi\nu_1^\prime)^{-1}\!\sim\!0.53$, except when the mass difference 
approaches its extremal values $\pm\nu_1^\prime\!=\!\pm(1\!-\!\nu_3)$. 
At these extremal values one of the $SU(3)$ constituents becomes massless 
and {\em delocalised}, which we excluded for $d\neq0$. 

Actually, the limit $d\!\rightarrow\!0$ is singular due to the {\em global} 
decomposition into $SU(2)\!\times\!U(1)$ at $d=0$. Gauge rotations $U$ in 
the global $SU(2)$ subgroup do not affect $\hat P_2$, and therefore any 
$UQ_1U^\dagger$ gives rise to the {\em same} accidental degeneracy. In 
particular solving $-3\pi\mu_2\!=\!\pi\nu_2^{\,\prime}\!-\!\half\partial_z
\acosh[\half\tr(\cA_2^{\prime} \cA_1^{\prime})]$ (corresponding to the Weyl 
reflection $Q_1\!\rightarrow\!Q_1^\dagger$) yields $z\!=\!9.868757$ for $\mu_2
\!=-\!1/30$ (isolated point in Fig.~\ref{fig:trace}). Indeed, 
$U\!\in\!SU(2)/U(1)$ traces out a (nearly spherical) {\em shell} where two 
eigenvalues of $P$ coincide (note that for $\mu_2\!=\!0$ this shell collapse 
to a single point, $z\!=\!10$). A perturbation tends to remove this accidental 
degeneracy. 

\section{Conclusions}

Topology is important for the non-perturbative understanding of non-abelian 
gauge theories, because it forces one to take serious that the configuration 
space is non-trivial~\cite{Ioffe}. That this has a role to play in the quantum 
theory can be easily understood from the difference between quantisation on 
the line versus the circle. 

Attempts to {\em localise} topology by gauge fixing (and its defects), should 
always be treated with healthy suspicion. This is illustrated by our 
interpretation of the knotted solitons in the Faddeev-Niemi model in terms 
of a non-linear maximal abelian gauge fixing of the gauge vacuum: There is 
nothing {\em physical} to be localised in a classical vacuum. Also we have 
shown that the notion of abelian projected monopoles has to be considered 
with care. For ``smooth" configurations and ``reasonable" choices of 
projections defects may well be correlated to action lumps, but this is 
by no means guaranteed. Furthermore, the defects can at best serve as bare 
quantities, which need to be dressed by the quantum corrections.

What is perhaps more important is that we learn that instantons and monopoles 
are closely connected, and should not be viewed in isolation. In disordered 
field configurations, gauge dislocations are likely to occur~\cite{Ioffe}. In 
four dimensions these are instantons (point dislocations), monopoles (line 
dislocations) and vortices (surface dislocations). Even when there is
a (sufficient) separation of scales, the problem remains how to isolate 
the effective low energy degrees of freedom in a reliable manner. 

\section*{Acknowledgements}
The first half of this talk is based on joint work with Andreas Wipf.
I thank him for collaboration, as well as for a provocative question 
that formed the inspiration for the the second half. I also thank Chris 
Ford for discussions and pointing out that the $SU(3)$ embedding of the 
$SU(2)$ caloron should give rise to defects forming a surface, rather 
than a point. Discussions with Poul Damgaard, Leander Dittmann, Gerald 
Dunne, Ludwig Faddeev, Thomas Heinzl, Marek Karliner, Martin L\"uscher, 
Holger Gies, Parameswaran Nair and Joaquin Sanchez Guillen are gratefully 
acknowledged, as well as hospitality at Cern. Finally I thank the 
organisers of the workshops in Cairns and Trento for their kind and
generous invitations.

\end{document}